\documentclass[twoside,leqno]{article}

\usepackage[letterpaper]{geometry}

\usepackage{siamproceedings}
\usepackage{wrapfig}
\usepackage[T1]{fontenc}
\usepackage{amsfonts}
\usepackage{graphicx}
\usepackage{epstopdf}
\usepackage{enumitem}
\usepackage{color}
\usepackage{algorithmic}
\usepackage[normalem]{ulem}

\ifpdf
  \DeclareGraphicsExtensions{.eps,.pdf,.png,.jpg}
\else
  \DeclareGraphicsExtensions{.eps}
\fi

\newsiamremark{remark}{Remark}
\newsiamremark{hypothesis}{Hypothesis}
\crefname{hypothesis}{Hypothesis}{Hypotheses}
\newsiamthm{claim}{Claim}

\usepackage{amsopn}

\begin{document}

\title{\Large AI-Powered Assistant for Long-Term Access to RHIC Knowledge}
    
\author{
Mohammad Atif \thanks{Computational Science Department, Brookhaven National Lab, Upton, NY (USA) \textsuperscript{\P}fmohammad@bnl.gov} \textsuperscript{\P} 
\and Vincent Garonne\thanks{Physics Department, Brookhaven National Laboratory, Upton, NY (USA) \textsuperscript{\S}jeromel@bnl.gov}
\and Eric Lancon\textsuperscript{\dag}
\and Jerome Lauret\textsuperscript{\dag \S}
\and Alexandr Prozorov\thanks{FNSPE, Czech Technical University, Prague, Czech Republic}
\and Michal Vranovsky\textsuperscript{\ddag}
}

%  (\email{fmohammad@bnl.gov}).
% (\email{vincent.garonne@bnl.gov}).
%  (\email{elancon@bnl.gov}).
%   (\email{jeromel@bnl.gov}).
%  ( \email{alexandr.prozorov@cvut.cz}).
%   (\email{mvranovsk@ssh.sdcc.bnl.gov}).
  
\date{}

\maketitle

% Copyright Statement
% When submitting your final paper to a SIAM proceedings, it is requested that you include
% the appropriate copyright in the footer of the paper.  The copyright added should be
% consistent with the copyright selected on the copyright form submitted with the paper.
% Please note that "20XX" should be changed to the year of the meeting.

% Default Copyright Statement
\fancyfoot[R]{\scriptsize{Copyright \textcopyright\ 20XX by SIAM\\
Unauthorized reproduction of this article is prohibited}}

% Depending on which copyright you agree to when you sign the copyright form, the copyright
% can be changed to one of the following after commenting out the default copyright statement
% above.

%\fancyfoot[R]{\scriptsize{Copyright \textcopyright\ 20XX\\
%Copyright for this paper is retained by authors}}

%\fancyfoot[R]{\scriptsize{Copyright \textcopyright\ 20XX\\
%Copyright retained by principal author's organization}}

%\pagenumbering{arabic}
%\setcounter{page}{1}%Leave this line commented out.

\begin{abstract} 
As the Relativistic Heavy Ion Collider (RHIC) at Brookhaven National Laboratory concludes 25 years of operation, preserving not only its vast data holdings ($\sim$1 ExaByte) but also the embedded scientific knowledge becomes a critical priority. 
The RHIC Data and Analysis Preservation Plan (DAPP) introduces an AI-powered assistant system that provides natural language access to documentation, workflows, and software, with the aim of supporting reproducibility, education, and future discovery. 
Built upon Large Language Models using Retrieval-Augmented Generation and the Model Context Protocol, this assistant indexes structured and unstructured content from RHIC experiments and enables domain-adapted interaction. 
We report on the deployment, computational performance, ongoing multi-experiment integration, and architectural features designed for a sustainable and explainable long-term AI access. 
Our experience illustrates how modern AI/ML tools can transform the usability and discoverability of scientific legacy data.
\end{abstract}

\section{Introduction.}
The Relativistic Heavy Ion Collider (RHIC) has operated since the year 2000, generating over 600 publications, and will end with close to an exabyte of data toward its end of life. RHIC has been shaping the nuclear physics community through its flagship experiments: STAR, PHENIX, and sPHENIX \cite{abdallah2022search,abdulhamid2024estimate,adam2020beam,adam2020bulk,STAR_2005gfr,ichimiya2009status,roland2021sphenix,tannenbaum2019highlights}. With final data collection concluding in 2025, the RHIC Data and Analysis Preservation Plan (DAPP) was launched to ensure that this irreplaceable legacy remains accessible and reusable for scientific and educational purposes. DAPP aims at preserving not only the raw data and published results, but the vast repository of tacit knowledge—experimental know-how, analysis techniques, detector-specific insights, and institutional memory—that typically resides only in the minds of researchers and scattered across internal documentation. This challenge is particularly acute given the decades-long timescales of the experiments, where today's graduate students may need to access and understand methodologies developed twenty years ago.

Among DAPP's key innovations is an AI-powered assistant that enables users to query preserved knowledge via natural language, dramatically lowering the barrier to reuse and understanding of complex datasets and workflows. 
Unlike web search or public-facing Large Language Models like ChatGPT, Gemini, which rely on publicly available data, our AI assistant can tap into a wealth of specific, highly relevant, trusted, and curated knowledge tailored to our experiments, collaborations, and internal research processes.
This paper presents a custom web scraper followed by architecture and initial deployment of the AI assistant within STAR, and its future as a generalizable framework for sustainable scientific data access.

% 3.3 Integration with Preservation Infrastructure
% The assistant is part of a broader ecosystem including:
% OpenData portal: For curated, public-access datasets.
% InvenioRDM repository: For publications, metadata, and DOIs.
% REANA workflow engine: For executing preserved analyses in containers.

% Once our home-grown ChatBot is integrated with in-house, proprietary information, it will become an indispensable tool for our scientific community.
% a scientific ChatBot designed to outperform general AI by providing accurate, relevant, and deep contextual answers from trusted scientific sources. 
% A home-grown chatbot offers personalized, secure access using proprietary data. 
% Our chatbot should provide scientifically or technically "weighted" answers, offering more in-depth information than typical publicly accessible tools designed for non-experts. 

\subsection*{Challenges}
\label{sec:motivation}

Building such a system requires addressing several critical technical and operational demands beyond conventional chatbots. The assistant must ingest and organize many PB of heterogeneous data while fusing semantic search with containerized execution environments, enabling users to retrieve archived results and reproduce analyses without reviving legacy software stacks.
Strict collaboration agreements necessitate fine-grained role- and group-aware access controls that respect data-security tiers (public, collaboration-restricted, and controlled) while honoring publication embargoes. To maintain credibility, the system continuously refreshes its knowledge base with real-time literature and facility updates, validated against technically demanding insider-level questions that routinely challenge open web tools.
Usability requirements drive customizable user profiles that adapt explanation depth, visualization style, and API integration for novices, seasoned analysts, and detector experts alike, with engagement metrics informing continuous improvement.
This paper presents the architecture and initial deployment of this AI assistant using STAR data as the first implementation, establishing a template for future experiments like the Electron-Ion Collider (EIC) where knowledge preservation could become an integral part of experimental design from its outset.

\section{Recursive Multi-Format Web-Content Extraction Framework}

\begin{wrapfigure}{r}{0.43\textwidth}
  \begin{center}
    \includegraphics[width=0.45\textwidth]{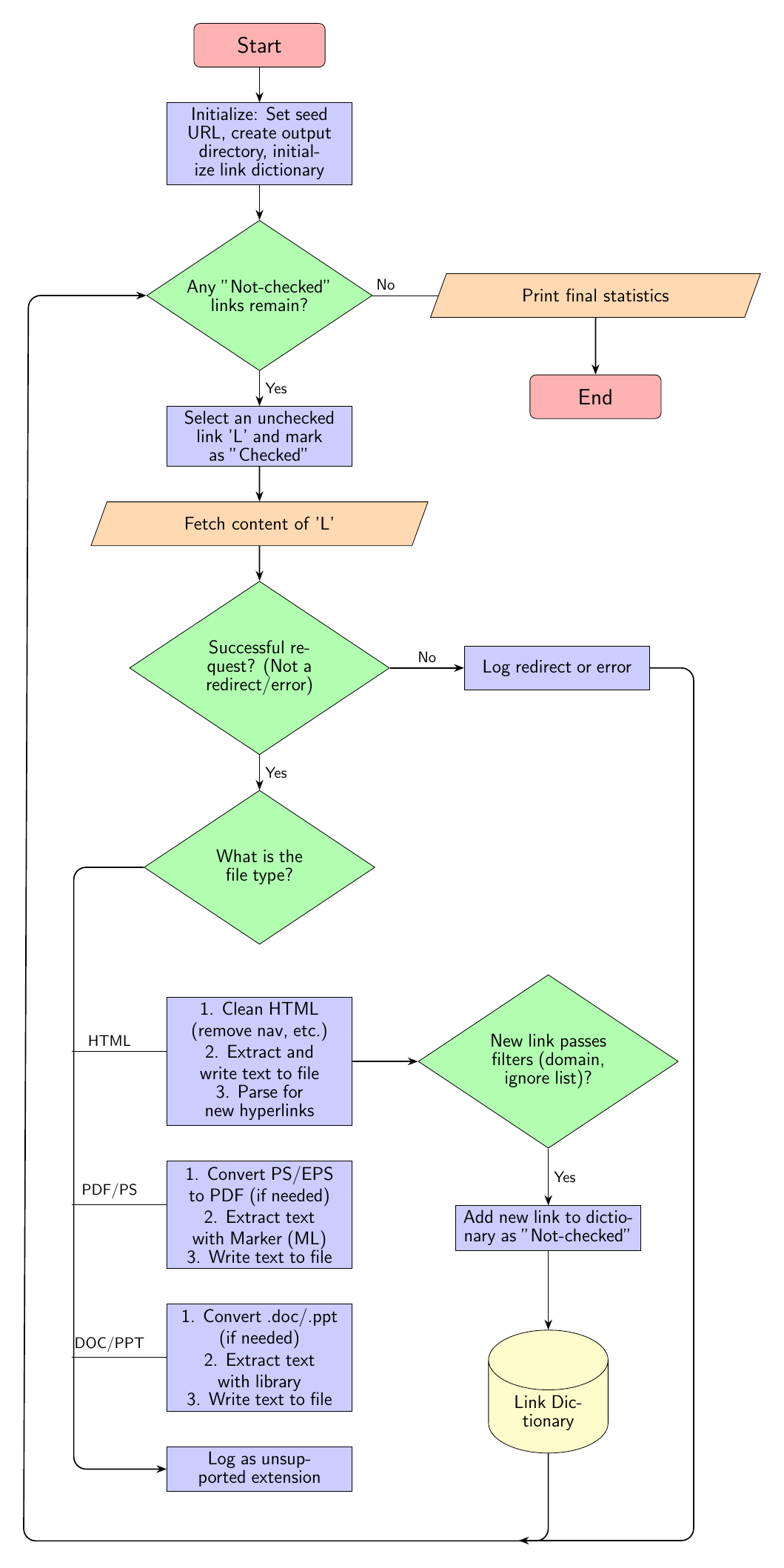}
  \end{center}
  \caption{Recursive Multi-Format Web-Content Extraction Framework}
  \label{fig:scraper}
\end{wrapfigure}

% **** We are missing a short intro speaking of the diverse elements?

% == Try 2

As foundational groundwork for building the AI assistant's knowledge base, we developed a comprehensive web content indexing system capable of systematically harvesting and processing diverse digital archives containing decades of experimental documentation, analysis notes, and institutional knowledge scattered across collaboration websites and document repositories. The bespoke multi-format content harvesting framework creates analysis-ready textual corpora from target web domains with deliberate focus on respectful rate-limiting and high-fidelity data extraction from academic and archival research environments (see Figure \ref{fig:scraper}).
The framework's architecture uses an iterative crawling mechanism starting from seed URLs, employing requests and BeautifulSoup \cite{BeautifulSoup} libraries to parse HTML pages and discover hyperlinks. A crucial "dynamic filter" prevents following external links, curating highly relevant, domain-specific datasets, while configurable blacklists exclude irrelevant content such as calendars and login pages.%—essential for maintaining signal-to-noise ratio.

Beyond standard HTML parsing, the system identifies linked documents by file extensions and invokes specialized parsers valuable for institutional repositories with mixed formats. For PDFs, it leverages the advanced marker library's machine learning models to convert visual structure into clean Markdown, significantly improving over traditional tools by preserving layout and semantic meaning \cite{marker_pdf}. The framework handles modern and legacy Microsoft Office documents (.docx, .doc, .pptx, .ppt) and PostScript files (.ps, .eps) through intelligent use of command-line utilities like LibreOffice and Ghostscript for on-the-fly conversions. Upon extraction, text undergoes rigorous cleaning and receives comprehensive metadata headers containing source URL, page title, and timestamp, ensuring clear data provenance. The framework reports detailed statistics on redirects, filtered links, and file extension frequencies, providing diagnostics that facilitate reproducibility studies and inform scraping optimization, thus creating a high-fidelity foundation for AI knowledge base construction.

\section{AI Assistant: Architecture and Deployment}

\subsection{Retrieval-Augmented Generation (RAG) and Model Context Protocol (MCP)}

The harvested content feeds into a Retrieval-Augmented Generation (RAG) architecture \cite{lewis2020retrieval} that embeds thousands of RHIC documents—technical notes, conference slides, code snippets, and software documentation—into a searchable ChromaDB vector database \cite{chromadb}. When users pose questions in natural language, the system retrieves semantically similar passages to inform responses with domain-specific context.
A critical architectural innovation is the Model Context Protocol (MCP) wrapper \cite{anthropic2024mcp,hou2025mcp}, which exposes each logical step of the assistant's reasoning chain—retrieval, summarization, inference, and evaluation—as composable "contexts" that can be independently configured and monitored. This protocol implementation via FastAPI-MCP decorates every `/ask` request with a lightweight JSON header specifying the ordered stack of models, maximum context budget per stage, and security tier inherited from the caller's session credentials.
At runtime, the MCP dispatcher constructs an orchestration graph capable of concurrent queries to multiple backends (inference engines or LLM APIs).
This is followed by attach relevant documents from ChromaDB and finally pipelining results to decoder LLMs for response generation. 
% The serializable nature of these contexts enables identical graph replay for offline regression testing or HPC batch processing without application code modification. 
This separation of execution logic ("what" to compute via the context graph) from deployment target ("where" to run on local GPU, cloud API, or containerized microservice) provides flexibility to incorporate emerging models while guaranteeing reproducibility, provenance tracking, and accountability across the entire workflow.

\subsection{Performance Comparison of Inference Engines}

% \begin{figure*}
%     \centering
%     \includegraphics[trim={0.5cm 0.35cm 1.2cm 1cm},clip,width=0.295\linewidth]{figures/throughput_A6000.png}
%     \includegraphics[trim={4cm 0.35cm 1.2cm 1cm},clip,width=0.23\linewidth]{figures/throughput_A100.png}
%     \includegraphics[trim={4cm 0.35cm 1.8cm 1cm},clip,width=0.345\linewidth]{figures/throughput_H100.png}    
%     \caption{Throughput (tokens/sec) observed on various inference engines across one, two, and four GPUs.}
%     \label{fig:throughput}
% \end{figure*}

% \begin{figure*}
%     \centering
%     \includegraphics[trim={0.5cm 0.35cm 1.2cm 1cm},clip,width=0.295\linewidth]{figures/utilization_A6000.png}
%     \includegraphics[trim={4cm 0.35cm 1.2cm 1cm},clip,width=0.23\linewidth]{figures/utilization_A100.png}
%     \includegraphics[trim={4cm 0.35cm 1.8cm 1cm},clip,width=0.345\linewidth]{figures/utilization_H100.png}    
%     \caption{Percentage utilization of GPUs on various inference engines across one, two, and four GPUs.}
%     \label{fig:utilization}
% \end{figure*}

\begin{figure*}
    \centering
    \includegraphics[trim={0.5cm 2.35cm 1.2cm 1cm},clip,width=0.295\linewidth]{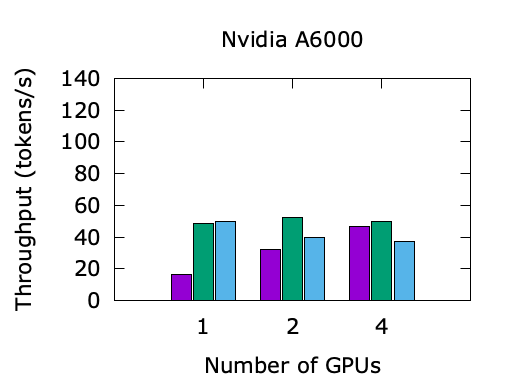}
    \includegraphics[trim={4cm 2.35cm 1.2cm 1cm},clip,width=0.23\linewidth]{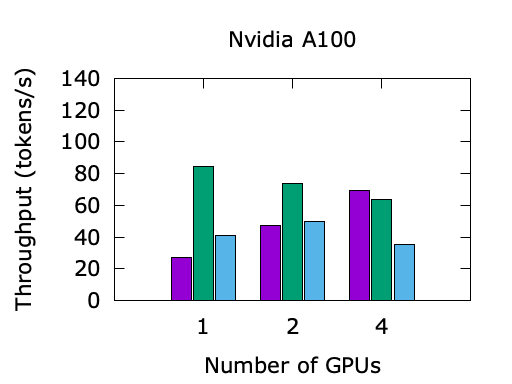}
    \includegraphics[trim={4cm 2.35cm 1.8cm 1cm},clip,width=0.345\linewidth]{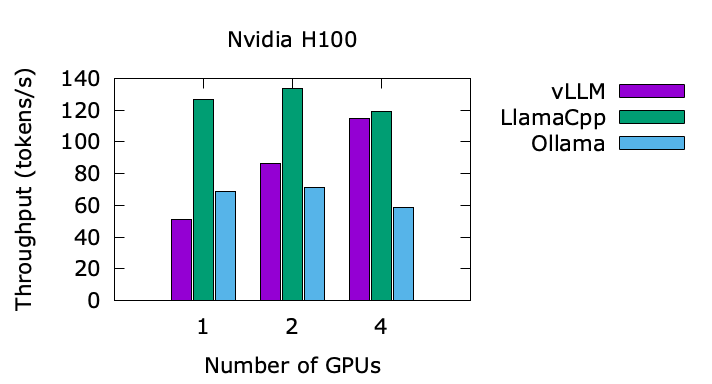}    
    \includegraphics[trim={0.5cm 0.35cm 1.2cm 2cm},clip,width=0.295\linewidth]{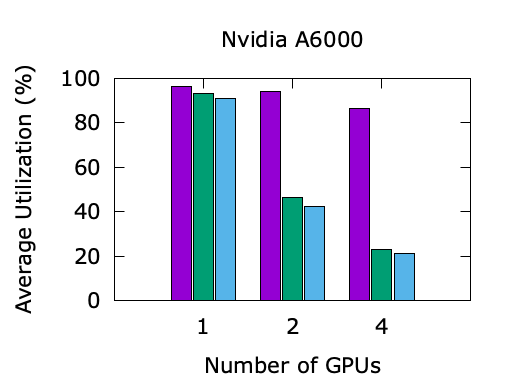}
    \includegraphics[trim={4cm 0.35cm 1.2cm 2cm},clip,width=0.23\linewidth]{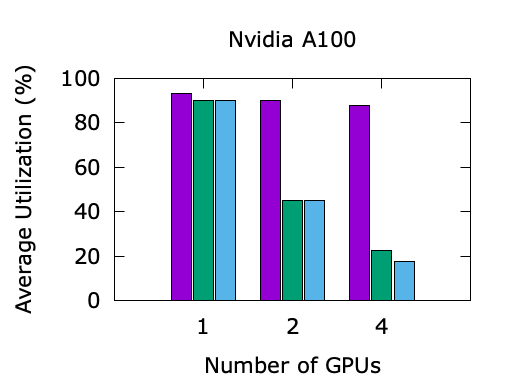}
    \includegraphics[trim={4cm 0.35cm 1.8cm 2cm},clip,width=0.345\linewidth]{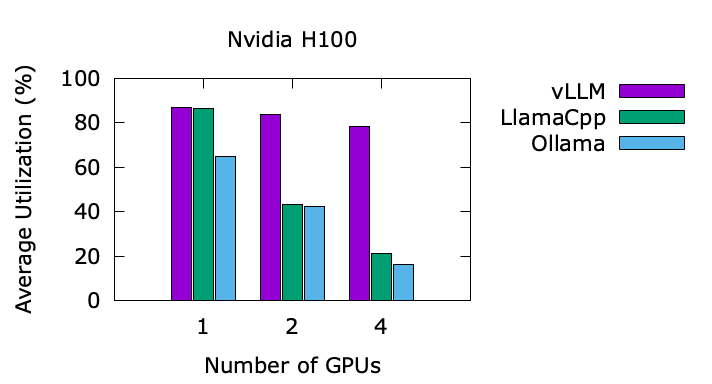}    
    \caption{Throughput (tokens/sec) (top row) and percentage utilization (bottom row) of GPUs on various inference engines across one, two, and four GPUs.}
    \label{fig:thruput_util}
\end{figure*}

Recently, several specialized LLM inference engines have been developed with different design goals for service-oriented infrastructures. Ref. \cite{park2025survey} surveyed 25 open-source and commercial inference engines for their optimization and performance characteristics. 
% They demonstrated that token-level throughput and time to respond hinge on an engine’s support for KV-cache reuse, continuous or hybrid batching, kernel fusion and low-bit quantization-optimizations.
The goal for a successful inference engine is to ameliorate GPU under-utilization while reducing per-request overheads.
Thus, a fair comparison must consider token throughput, GPU utilization, and the available hardware, because the very same model can exhibit markedly different behavior across engines and deployments.
The remainder of this section therefore compares three representative engines (vLLM, LlamaCpp, Ollama) to illuminate the practical trade-offs among speed, scalability and cost that shape modern chatbot services.

The analysis of inference engine performance across different GPU architectures in Figure \ref{fig:thruput_util} reveals distinct trends in throughput and resource utilization, heavily influenced by both the hardware generation and the underlying design philosophy of the software. As expected, throughput scales directly with the generational advancements of the GPU hardware; for any given engine and configuration, the newer H100 consistently outperforms the A100, which in turn surpasses the older A6000 architecture. However, the scaling observed on increasing the number of GPUs highlights fundamental differences in the engines' design philosophies. vLLM, for instance, demonstrates a clear design focus on maximizing throughput via parallelization, as its tokens-per-second rate consistently increases with the addition of GPUs while maintaining high average utilization, typically above 80\%. In stark contrast, LlamaCpp, while often exhibiting the highest single-GPU throughput, shows a dramatic decrease in resource utilization as more GPUs are added, dropping from ~90\% on a single GPU to as low as ~21\% on four. This suggests its multi-GPU strategy may prioritize memory distribution for larger models over computational parallelism, leading to under-utilization of compute resources on additional GPUs. Ollama's performance scaling is generally less efficient than the other two, with both throughput and utilization declining as the GPU count increases. In summary, while newer hardware like the H100 provides the highest raw performance ceiling, achieving optimal multi-GPU throughput is critically dependent on the inference engine's architecture, with vLLM's parallel-first approach showing superior scaling over LlamaCpp's apparent memory-centric distribution model in these tests.
  
\section{Benchmarking, Validation, and Metrics of Success}
% Validation Framework: Incorporates expert feedback and usage metrics to refine responses and filter hallucinations.

% Technology Watch: Annual reviews ensure compatibility with evolving formats, AI methods, and hardware platforms.

% FAIR Alignment: All outputs are designed to be Findable, Accessible, Interoperable, and Reusable.
It is important to define success metrics for a specialized scientific ChatBot as many generic LLMs (ChatGPT, Gemini, Claude, ...) exist and available to end-user, challenging the merits of an in-house tool. Our approach is tailored to the scientific community it is focused on our ability to deliver accurate, contextually deep answers from trusted scientific sources, surpassing general AI tools. Key criteria include precision, seamless integration with scientific knowledge databases, enhanced time efficiency, and comprehensive search capabilities across diverse scientific domains. In addition, we will provide secure access to proprietary data, while addressing the challenge of keeping it updated with real-time external information through strategies like automated data synchronization and hybrid search approaches (where both real-time and local database are consulted and the results merged).

% \sout{In our approach, model responses were systematically evaluated against expert-validated reference answers using a multi-faceted assessment approach. Domain specialists with many years of experience established ground truth answers for each benchmark question, providing the basis for substantial inter-rater reliability. Significance of performance differences between RAG-enhanced and baseline models was observed. Our initial (manual, expert vetted) evaluation revealed that models incorporating RAG knowledge consistently outperformed baseline variants, with embedding-space similarity analysis identified as a promising direction for future evaluation enhancement.}

In this work, we qualitatively evaluate LLM responses against expert-validated reference answers.
Domain specialists with several years of experience have established ground truth answers for each benchmark question like "How are space charge effects considered in the STAR experiment?" and "What is the TOF resolution in STAR experiment?"
We compare the answers to the above question generated by Llama3.3-70B, Mistral-Large-2411, both augmented with the
RAG framework accessing local documents, and a commercial version of ChatGPT o3.
The evaluation revealed distinct performance characteristics. 
For an initial query all three models provided relevant and accurate information, with stylistic variations. 
The Llama3.3 model was noted for its concise and detailed narrative, whereas, Mistral was similarly precise but slightly more verbose. In contrast, ChatGPT o3 adopted a more pedagogical, verbose style that, while comprehensive, included extraneous details.% that might be extraneous for an initial expert query.

A key aspect of the comparison was the ability to cite sources. While the RAG-based models inherently provide references from their local knowledge base, it was observed that ChatGPT o3 also cited its sources, which included public archives like arXiv, academic networks such as ResearchGate, and even publicly accessible files on institutional websites.
The primary differentiator and principal value of the local RAG-based approach emerged from its ability to incorporate private, unpublished information while providing excellent answers at minimal operational costs. Specifically, Llama3.3 and Mistral models successfully and quickly retrieved context from internal collaboration mailing lists, which are inaccessible to public-facing commercial LLMs. This access to informal scientific discourse—including technical discussions, troubleshooting exchanges, and expert clarifications from mailing lists and conversations—may explain why the RAG-based responses appeared more aligned with the practical, conversational tone that scientists expect when consulting colleagues. Therefore, we conclude a locally deployed RAG system holds a distinct and critical advantage for queries that depend on proprietary or non-public data. We further speculate that in cases where information is not yet published, a RAG-based approach leveraging internal documents is not just valuable but essential for comprehensive and accurate scientific inquiry. The development of a formal, multi-dimensional evaluation benchmark to systematically quantify these observations is currently in progress.
  
% On the more specific query related to Time-of-Flight (TOF), a significant performance divergence was observed. ChatGPT o3 markedly outperformed the RAG-based models, delivering a more thorough and historically nuanced response that included year-by-year variations. This result was notably different from previous informal tests, suggesting rapid evolution in the capabilities of commercial models.

\section{Conclusion and future work}

% --- Jerome's replacement, expanding with some of our discussed plans

We presented a domain-specific AI assistant that effectively preserves and serves nuclear physics knowledge through RAG architecture enhanced with MCP orchestration. The system's key advantage lies in accessing proprietary data sources—including internal mailing lists and unpublished materials—unavailable to commercial LLMs. Initial evaluation confirms RAG-enhanced models deliver contextually appropriate responses aligned with scientific expectations, demonstrating essential value for knowledge preservation in collaborative research environments.
Our evaluation methodology established expert-validated reference answers with substantial inter-rater reliability, revealing significant differences between RAG-enhanced and baseline models. 
% This manual, expert-vetted approach demonstrated consistent RAG superiority, with embedding-space similarity analysis identified as a promising evaluation enhancement.
Future work will deploy independent LLM judges using structured prompts and explicit evaluation criteria to assess outputs across factual accuracy, completeness, relevance, and clarity dimensions.

Our immediate plans focus on extending the AI assistant to all RHIC experiments, implementing a comprehensive role-aware access control fabric that cleanly separates public, collaboration-restricted, and controlled artifacts. Security implementations will address the sensitive nature of proprietary experimental data through enterprise-grade encryption (AES-256 at rest, TLS 1.3 in transit), institutional SSO integration, and comprehensive audit logging. The role-aware access control system maps user authentication tokens to document classification hierarchies, enabling permissions that respect collaboration agreements.
% while maintaining data sovereignty requirements essential for international research partnerships. 
Custom ingestion pipelines will incorporate community-adopted binary formats alongside heterogeneous media, capturing tacit expertise before it is lost to staff turnover. 
The web scraping framework requires enhancement for scalability, fault tolerance, and authentication support for protected institutional resources, with asynchronous I/O improving throughput while maintaining respectful access patterns. 
% Persistent frontier management through SQLite or Redis will enable distributed crawling and fault recovery, complemented by a plugin registry for extensible format support.

To maintain scientific currency, we plan an adaptive synchronization layer combining automated harvesting of open-access publications via APIs and RSS feeds, on-demand trusted web searches for queries beyond the local corpus, and human-in-the-loop dashboards enabling domain specialists to flag and annotate emerging results. The framework's domain-agnostic modularity positions it as a knowledge-stewardship engine for the forthcoming Electron-Ion Collider era and beyond, 
% with iterative search logging and benchmark-guided optimization driving continuous improvement toward 
thus leading to a sustainable, self-evolving assistant that integrates reproducibility, education, and discovery at scale.

\bibliographystyle{siamplain}
\bibliography{references}

\end{document}